\documentclass[notitlepage,10pt,
onecolumn,amsmath,amssymb
]{revtex4-1}
\def\d{{\rm d}}
\def\e{{\rm e}}
\def\i{\ifmmode{\rm i}\else\char"10\fi}
\def\I{{\rm I}}
\newcommand{\nz}{\mathbb{N}}
\newcommand{\rz}{\mathbb{R}}
\newcommand{\cz}{\mathbb{C}}

\begin{document}
%\begin{opening}
\title{PATH INTEGRALS AND LIE GROUPS\footnote{Published in: E.A.\ Tanner and R.\ Wilson (eds.), {\it Noncompact Lie Groups and Some of Their Applications}, (NATO ASI Series, Kluwer Academic Publishers, Dordrecht, 1994) p.\ 199.}}
\author{Akira Inomata}
\affiliation{Department of Physics, State University of New York at
Albany,\\
Albany, N.Y.\ 12222, USA}
\author{Georg Junker}
\affiliation{Institut f\"ur Theoretische Physik I, Universit\"at
Erlangen--N\"urnberg, Staudtstr.~7, D--91058 Erlangen, Germany.}
%\end{opening}

%\begin{document}
\begin{abstract}
The roles of Lie groups in Feynman's path integrals in non-relativistic
quantum mechanics are discussed. Dynamical as well as geometrical
symmetries are found useful for path integral quantization. Two examples
having the symmetry of a non-compact Lie group are considered. The first
is the free quantum motion of a particle on a space of constant negative
curvature. The system has a group $SO(d,1)$ associated with the
geometrical structure, to which the technique of harmonic analysis on a
homogeneous space is applied. As an example of a system having a
non-compact dynamical symmetry, the $d$-dimensional harmonic oscillator
is chosen, which has the non-compact dynamical group $SU(1,1)$ besides
its geometrical symmetry $SO(d)$. The radial path integral is seen as a
convolution of the matrix functions of a compact group element of
$SU(1,1)$ on the continuous basis.
\end{abstract}

\maketitle
%Introduction
%\input{ctexas1.tex}
%\input{tx1.tex}
\section{Prologue}
Needless to say,  Lie groups have a special importance in quantum
mechanics. The Lie group often details the kinematical symmetry of a
quantum system. The analysis of angular momentum on the basis of the
rotation group $SO(3)$ or the unitary group $SU(2)$ is a classic
example. The Lie group sometimes reveals itself in the structure of
quantum dynamics. The well-known accidental degeneracy of the Coulomb
problem has been ascribed to its dynamical symmetry $SO(4)$ which is not
shared by other spherically symmetric systems. In the applications to
non-relativistic quantum mechanics, it is usually associated with
Schr\"odinger's differential equation or involved in making a algebraic
framework of a quantum system. Rather independent of physics, there are
extensive studies of differential equations and special functions from
the Lie group representation aspect \cite{Mi68,V68}. The
spectrum-generating algebraic approach has placed many of the standard exactly
soluble problems in quantum mechanics within the $su(1,1)$ algebraic
scheme and opened the field of infinite component theories for composite
systems \cite{B71}. However, little is known about the roles of Lie
groups in Feynman's path integral. As the kernel of the heat equation is
easily obtainable from the Fourier analysis, the path integral
representing the kernel of the Schr\"odinger equation may be studied
from a more general harmonic analysis point of view. In recent years,
there have been considerable developments in the study of the Lie group
theoretical approach to Feynman's path integral [4-10]. In this report, we
would like to highlight the main ideas and results of the study.

Any standard text book of quantum mechanics contains the treatment of
central potential problems in polar coordinates, whereas nearly all of
the books dealing with Feynman's path integral ignore the polar
coordinate treatment. This is simply a reflection of the situations that
path integration is difficult in non-cartesian variables and hence that
Feynman's path integral can be explicitly calculated only for the
harmonic oscillator or more general quadratic systems. The polar
coordinate formulation was already initiated, to the authors' knowledge, by
Ozaki \cite{Oz55} as early as 1955, and independently, by Edwards and
Gulyaev \cite{EG64} in 1964. They separated the angular part and the
radial part of the path integral for a spherically symmetric system in
polar coordinates and  calculated the propagator for a free particle.
The angular path integration for a free particle or a particle in a
central potential is in essence an application of harmonic analysis
based on the unitary representation of $SO(3)$ -- an elementary
application of the Peter-Weyl theorem.

However, the method of explicit
calculation of the radial path integral \cite{PI69} was not known
until 1969, without which the polar coordinate path integral has of
little interest. The energy spectrum of a central potential problem
arises only from the radial path integral. Harmonic analysis of the
radial path integral is not as simple as that of the angular part.
For a free particle, extension of $SO(3)$ to the Euclidean group $E(3)$
is sufficient for covering both the radial part as well as the angular
part \cite{BJ89}. In the presence of a central potential, the
translational symmetry is broken, so that the Euclidean group becomes
inapplicable. It turns out that the radial path integral is most
conveniently analyzed on the basis of the dynamical group $SU(1,1)$
\cite{I92}.

In 1968, in order to demonstrate that the sum over classical paths is
exact, Schulman \cite{S68} calculated semiclassically the path integral
for the symmetric top on the manifold of $SU(2)$. In relation with
Schulman's observation, Dowker \cite{D71} showed that the sum over
classical paths is exact on the manifold of a class of Lie groups. Since
explicit evaluation of Feynman's path integral was difficult, attention
was mainly focused on semiclassical approximation and examination of its
exactness. What Schulman and Dowker have studied are harmonic expansions of the
propagators in terms of spherical functions. The semiclassical
approximation is not generally exact even if space is symmetric.

It is a strange fact that Feynman's path integral cannot provide the
solution for the hydrogen atom problem. Since no closed form expression
for the propagator of the hydrogen atom is available, there have been a
number of attempts to construct one. Historically, Feynman's path
integral was once expected as a possible means to find the Coulomb
propagator. Soon it was recognized that the power of the path integral
method was very limited in deriving exact results. None of the integral
representations of the Coulomb propagator so far available was
originated from Feynman's path integral. In the summers of 1976-78, the
Kustaanheimo-Stiefel transformation became a topic of discussions in the
seminars of the University of Munich. Barut and Wilson \cite{BSW79} were mainly
investigating the KS mapping as a realization of the subgroup of a wider
dynamical group $SO(4,2)$. One of the present authors examined its use
in the path integral for the hydrogen atom, and realized that the path
integral given in the KS coordinates does not help in constructing a closed
form expression for the Coulomb propagator. The discussions of Munich were
carried over to Trieste in 1978, involving Barut, Duru, Wilson and
others \cite{BDW}. Barut was optimistic about the use of the KS transformation in
path integration. In 1979, Duru and Kleinert \cite{DK79}, formally
applying the Kustaanheimo-Stiefel transformation of space and time to
the Hamiltonian path integral for the hydrogen atom, succeeded to derive an
integral representation for the Coulomb propagator. Giving up the hope
for expressing the Coulomb propagator in closed form, the authors of
ref.\ [\onlinecite{HI82}] applied the KS transformation to the Lagrangian path
integral, carrying out the explicit time-sliced path integral
calculation, to arrive at the exact expression for the energy-dependent
Green function of the hydrogen atom, previously obtained by Hostler
\cite{H64} from Schr\"odinger's equation.

The Kustaanheimo-Stiefel transformation was originally introduced for
regularization of the classical Kepler orbit \cite{KS65}. It consists of
a space transformation $\rz ^{3} \to \rz ^{4}$ and a position-dependent
time transformation. In particular, the time transformation is
integrable only along the classical orbit. In path integration, there is
no unique orbit, and such a formal calculation often involves ambiguity \cite{I86}.
The significance of these results is that Feynman's path integral, if
slightly modified, can produce solutions for systems other than
quadratic systems. Furthermore, the successful use of the
Kustaanheimo-Stiefel map that takes the geometrical symmetry group
$SO(3)$ to the dynamical symmetry group $SO(4)$ suggests that the
dynamical symmetry could play a role in path integral calculation. The
fact of matter is that the path integral for the Coulomb problem can be
solved in terms of polar variables without the help of the KS
coordinates \cite{I84}. Yet, the local time transformation of Kustaanheimo
and Stiefel was found essential for solving the Coulomb path integral
problem.

Noticing that the P\"oschl-Teller oscillator has the dynamical
symmetry of $SU(2)$ [\onlinecite{BIW87a}], the one-dimensional path integral for this system
is extended to the path integral on $SU(2)$ with the aid of the
asymptotic form of an integral representation for the modified Bessel
function \cite{IK85,IW86}. This nontrivial dimensional extension of the
path integral has enabled us to carry out path integration in the way
angular path integration was completed. Naturally, a particle bound in
the modified P\"oschl-Teller potential, having the dynamical symmetry of
$SU(1,1)$ [\onlinecite{BIW87b}], can be path integrated in $SU(1,1)$.
In fact, in combination
with a local time transformation of the Kustaanheimo-Stiefel type, a
number of systems have been solved in the mani\-fold of $SU(2)$ and
$SU(1,1)$. Of course, $SU(2)$ and $SU(1,1)$ can be identified with
$S^{3}=SO(4)/SO(3)$ and $\Lambda ^{3}=SO(3,1)/SO(2,1)$, respectively. We
shall call those systems path integrable on $S^{d}$ or $\Lambda ^{d}$
the systems of the {\em hypergeometric class}. The stochastic processes
corresponding to Feynman's path integrals of the hypergeometric class
are called the {\em Legendre processes} \cite{FLM92}.
Although a system of $SU(2)$ can be
converted to a system of $SU(1,1)$ by analytic continuation, the role of
the non-compact group $SU(1,1)$ in path integration is not completely
expressed by analogy from the case of $SU(2)$. More subtle
treatments are needed for non-compact systems. On these backgrounds, a
general formulation of the path integration on a group manifold, compact
or non-compact, has been developed \cite{BJ87,J89b}.

As has been mentioned above, the radial path integral can be handled in
a way very similar to the spectrum generating algebra of $SU(1,1)$.
Although this method yields solutions only for an isotropic harmonic
oscillator in an inverse-square potential, the path integral for a
number of systems can be transformed into such a standard harmonic
oscillator form if the local time transformation technique is used
\cite{I92}. We shall refer to the systems soluble by radial path
integration as those of the {\em confluent hypergeometric class}. Again,
the stochastic counterparts are called the {\em Bessel processes}
\cite{FLM93}. For
the hypergeometric class, the spherical functions involved are the
matrix elements on the discrete basis. In the case of the confluent
hypergeometric class, the matrix elements adopted are on a continuous
basis \cite{IJ93}. The matrix elements of $SU(1,1)$ on continuous bases
have been discussed by Barut and Fronsdal \cite{BF66}, Barut and Phillips
\cite{BP68}, Lindblad and Nagel \cite{LN70}, and Mukunda and Radhakrishnan
\cite{MR73}.

Section 2 briefly reviews the relations of Feynman's path integral to
the evolution operator, the propagator, the resolvent, the
energy-dependent Green function and the promotor. In Section 3, path
integrals on homogeneous spaces are discussed. In Sections 4 and 5,
two specific path integrals are solved with the help of harmonic analysis.
Section 4 deals with a free particle on a space of constant negative
curvature and Section 5 analyses the dynamical group of the path
integral for the harmonic oscillator in $d$-dimensions. A brief review
of the rudiments of harmonic analysis is given in Appendix.

\section{Feynman's Path Integral}
Let us start with the Hamiltonian of the standard form, 
\begin{equation}
H=H_{0}+V({\bf x})={\bf p}^{2}/(2M)+V({\bf x}),
\end{equation}
which acts on the Hilbert space ${\cal H}={\cal 
L}^{2}(\rz^{d})$. The operators ${\bf p}$ and ${\bf x}$ represent the 
momentum and position of a particle of mass $M$ moving in the 
$d$-dimensional Euclidean space $\rz^{d}$ under the influence of a 
scalar potential $V$. They obey Heisenberg's commutation relations 
$x_{i}p_{k} - p_{k}x_{i}=\i\hbar\delta _{ik}$ $(i,k=1,2,\ldots, d)$ where 
$x_{i}$ and $p_{k}$ are cartesian components of ${\bf x}$ and ${\bf p}$. 
As the Schr\"odinger operator acting on the Hilbert space ${\cal 
H}={\cal L}^{2}(\rz^{d})$, the Hamiltonian (1) is self-adjoint with 
$D(H)=C^{\infty }_{0}(\rz^{d})$, but not necessarily bounded. The time 
evolution of the system from the initial time $t_{0}$ to the final time 
$t$ is described by the unitary evolution operator $U(t,t_{0})=\exp[-(\i/\hbar)(t-t_{0})H].$
As long as the Hamiltonian (1) is self-adjoint, the time evolution 
forms a one-parameter group. Under the causal restriction $t \geq t_{1} 
\geq t_{0}$, the evolution operator satisfies the initial condition,
$U(t_{0}, t_{0})=1,$ and the composition rule, $ 
U(t, t_{1})U(t_{1}, t_{0})=U(t,t_{0}).$

The propagator is a matrix element of the evolution operator 
in the position representation defined only for $t''>t'$:
\begin{equation}
K({\bf x}'', {\bf x}'; t''- t')=\langle {\bf x}'' \mid 
\exp\{-(\i/\hbar)(t''-t')H\}\mid {\bf x}'\rangle,
\end{equation} 
where ${\bf x}'={\bf x}(t')$ and ${\bf x}''={\bf x}(t'')$. The initial 
condition of the evolution operator provides us the normalization 
condition, 
\begin{equation}
\lim_{t \rightarrow t'}\, K({\bf x}, {\bf x}'; t- t')=\delta({\bf x}-{\bf x}').
\end{equation}
From the composition rule readily follows the semi-group property,
\begin{equation}
K({\bf x}'', {\bf x}'; t''- t')=\int\limits_{\rz^{d}} \d {\bf x}\, 
K({\bf x}'', {\bf x}; t''- t)\, K({\bf x}, {\bf x}'; t- t'),
\end{equation}
where d${\bf x}$ is the translation-invariant Lebesgue measure in 
$\rz^{d}$. The property (4) leads to 
\begin{equation}
K({\bf x}'', {\bf x}'; t''-t')=
\int\limits_{\rz^{d}} \d {\bf x}_{1} \cdots \int\limits_{\rz^{d}} \d 
{\bf x}_{N-1} \, K({\bf x}_{N}, {\bf x}_{N-1}; \tau _{N})\cdots 
K({\bf x}_{1}, {\bf x}_{0}; \tau _{1}),
\end{equation}
where ${\bf x}_{j}={\bf x}(t_{j})$, $\tau _{j} = t_{j} - t_{j-1} > 0$, 
$t'=t_{0}$ and $t''=t_{N}$. For convenience, as Feynman did, we adopt 
now on the isometric subdivision of the time interval $\tau =t''-t'$
as $\tau _{j}=\tau /N = \epsilon $ for all $j$. 

In the celebrated 1948 paper, 
Feynman \cite{F48} asserted that the (infinitesimally) short-time propagator 
can be given by   
\begin{equation}
\tilde{K}({\bf x}_{j}, {\bf x}_{j-1}; \epsilon )=\left[\frac{M}{2\pi\i\hbar 
\epsilon }\right]^{d/2}\,
\exp\left[\frac{\i}{\hbar}S_{\epsilon }({\bf x}_{j}, {\bf x}_{j-1})\right],
\end{equation}
with the short-time action,
\begin{equation}
S_{\epsilon }({\bf x}_{j}, {\bf x}_{j-1})=\frac{M}{2\epsilon }
\left({\bf x}_{j} - 
{\bf x}_{j-1}\right)^{2} - \frac{1}{2}\epsilon  \Bigl(V({\bf x}_{j}) + 
V({\bf x}_{j-1})\Bigr). 
\end{equation}
Feynman's assertion implies that the propagator can be calculated with 
(6) by the infinite convolution formula, 
\begin{equation}
K({\bf x}'', {\bf x}'; \tau )=\lim_{N \rightarrow \infty } 
\int\limits_{\rz^{d}}\,\prod_{j=1}^{N-1} \d {\bf x}_{j} 
\,\prod_{j=1}^{N}\,\tilde{K}({\bf x}_{j}, {\bf x}_{j-1}; \epsilon ).
\end{equation} 
Feynman's formula for the propagator was proven 
stochastically by Kac \cite{Kac59} for an imaginary time $\beta = 
\i\tau/\hbar >0$. For real time, Nelson \cite{Nel64} has a proof for 
potentials of the Kato class, and Faris \cite{Far67,Sim71} has a proof 
for the Rollnik class. Both are based on Trotter's product formula, 
\begin{equation}
\e^{(\i/\hbar)\tau H} = \lim_{\varepsilon \downarrow 0}\lim_{N 
\rightarrow \infty }\left(\e^{(\i/\hbar)\tau H_{0}/N(1 - \i\varepsilon 
)}\,\e^{(\i/\hbar)\tau V/N}\right)^{N}. 
\end{equation} 

In practice, Feynman's path integral can be evaluated only for quadratic 
systems.
To expand the scope of Feynman's 
path integral, we also pay attention to the resolvent of the Hamiltonian, 
\begin{equation}
G=\frac{1}{E - H} = \frac{1}{\i\hbar}\int\limits_{0}^{\infty 
}\d\tau \,\exp\{(\i/\hbar)\tau (E-H)\}, ~~~~~~\mbox{Im}\,E > 0.
\end{equation}
The matrix element of the resolvent in the position representation, which 
is often referred to as the energy-dependent Green function, is given by 
\begin{equation}
G({\bf x}'', {\bf x}'; E)=\langle {\bf x}''|(E-H)^{-1}|{\bf x}' \rangle 
=\frac{1}{\i\hbar}\int\limits_{0}^{\infty }\d\tau\,P({\bf x}'', {\bf 
x}';\tau),
\end{equation}
where 
\begin{equation}
P({\bf x}'', {\bf x}'; \tau )= \langle {\bf x}''|\exp\{(\i/\hbar)\tau (E-
H)\}|{\bf x}' \rangle .         
\end{equation}
The last entity $P({\bf x}'', {\bf x}'; \tau )$, which we call the {\it promotor}, is also expressible 
as a path integral in Feynman's form. The Green function contains the 
same quantum-mechanical information as that the propagator has. 
Therefore, the path integral may be calculated for the promotor rather 
than the propagator. Once the Green function is found with the help of 
the promotor, it can also be converted into the propagator by a Fourier 
transformation. 

It is true that the path integral structure of the promotor is identical 
with that of the propagator except for the additional energy term. 
When Feynman's path integral is difficult to evaluate, the 
path integral for the promotor is equally difficult to calculate. 
Nevertheless, the promotor has a unique advantage. Suppose it is 
transformed into 
\begin{equation}
\tilde{P}({\bf x}'', {\bf x}';\sigma )= \langle {\bf 
x}''|\exp\{(\i/\hbar)\sigma f({\bf x})(E - H)g({\bf x})\}|{\bf x}' \rangle 
\end{equation}
with $\sigma = \tau /[f({\bf x}')g({\bf x}'')]$ and positive-definite 
$q$-number functions $f({\bf x})$ and $g({\bf x})$. Then we can show 
that the Green function (11) can also be evaluated by 
\begin{equation}
G({\bf x}'', {\bf x}'; E)
=\frac{1}{\i\hbar}\int\limits_{0}^{\infty }\d\sigma\,\tilde{P}({\bf x}'', {\bf 
x}'; \sigma )\,(\d\tau/\d\sigma ) .
\end{equation}
The result remains the same. This is a remarkable property. The above
time transformation of the promotor is sometimes called 
the {\em local time rescaling trick} \cite{I92}. This added flexibility 
in the path integral has contributed much to the development of path 
integral calculus in the last decade. In this regard, when we talk about 
Feynman's path integral, we may include the path integral for the 
promotor as well as the original path integral for the propagator. All 
the arguments given now on are equally applicable to the propagator and 
the promotor.

Suppose Feynman's assertion is fully justified for a certain class of 
potentials. However, we have to note that the background space 
of the time-sliced path integral (11) is $\rz^{d}$. In fact, Feynman 
recognized that the cartesian coordinate system played a special 
role in defining Feynman's path integral, and suggested that 
only the short-time propagator found in cartesian variables may be 
expressed by a coordinate transformation in any choice of 
coordinate variables \cite{F48}. The special role of the cartesian 
coordinates is not unique in Feynman's path integral. We can find a 
similar situation in the canonical quantization procedure.
Recall that Heisenberg's commutation relations are applicable only 
to cartesian variables. In setting up Schr\"odinger's equation from the 
Hamiltonian (1), we have to replace ${\bf p}^{2}$ by $-(\i/\hbar)^2\nabla 
^{2}$ in cartesian variables. Then, for a spherically symmetric system, 
for instance, we transform the Lapalace-Beltrami operator $\nabla^{2}$ 
from cartesian variables to polar coordinates. The connection of these 
two similar situations must be of a profound significance.
  
By expressing the short-time propagator 
in polar coordinates and expanding in terms of Legendre functions 
(the zonal spherical functions on $S^{2}$), Feynman's path integral in 
three dimensions has been separated into the radial path integral and 
the angular part \cite{EG64,PI69} as 
($r'=|{\bf x}'|$, $r''=|{\bf x}''|$)
\begin{equation}
K({\bf x}'',{\bf x}';\tau )=\frac{1}{4\pi }\sum_{l=0}^{\infty}(2l + 
1)\,K_{l}(r'',r';\tau 
)\,P_{l}({\bf x}'\cdot {\bf x}''/r'r'').
\end{equation}
Similarly, it is possible to express Feynman's path integral in any 
desired coordinate system as long as the background space remains flat. 
However, there is no established formulation of a path integral in a 
general curved space. In this article, our interest is not in 
formulating a path integral in a general curved space. We are rather 
interested in the questions as to how we can take advantage of a 
symmetry of the physical system when we carry out path integration 
and whether the path integral can be extended to a homogeneous space (or 
more restrictively a symmetric space) and solved in much the same 
fashion that the polar coordinate path integral is treated.

\section{\bf Path Integral on a Homogeneous Space}
Let a group $G$ be a transformation group on a space ${\cal M}$.
If $G$ acts transitively on ${\cal M}$,  then ${\cal M}$ is a 
homogeneous space with respect to $G$. If $H$ is the isotropy group of 
$G$ at a point $q_{a}$ of ${\cal M}$, then ${\cal M}=G/H$. 
For details, see Appendix. 

In an effort to understand Feynman's path integral on a homogeneous 
space, we assert that the propagator on a space ${\cal M}$ equipped with 
a measure $\d q$ can be given by the following multi-convolution, 
($q''=q_{N}$, $q'=q_{0}$)
\begin{equation} 
K(q'', q';\tau )=\lim_{N\to\infty } 
\int\limits_{\cal M}^{~}\d q_{1}\cdots\int\limits_{\cal M}^{~}\d q_{N-1}\, 
\tilde{K}(q_{N},q_{N-1};\epsilon )\cdots \tilde{K}(q_{1},q_{0};\epsilon ),  
\end{equation}
in analogy to the time-sliced path integral (8). The 
finite-time propagator in 
${\cal M}$ is defined in the Hilbert space ${\cal H}={\cal L}^{2}({\cal 
M})$. Here we assume that it has the following properties,
\begin{equation}
\lim_{t \rightarrow t'}K(q,q';t-t')=\delta (q-q'),
\end{equation}
\begin{equation}
\int\limits_{\cal M}^{~} \d q \,K(q'',q;t''-t)\,K(q,q';t-t') = 
K(q'',q';t''-t).
\end{equation}
Note that the finite-time propagator $K(q'', q';\tau )$ approaches the 
short-time propagator $\tilde{K}(q, q';\epsilon )$ as $\tau $ tends to 
$\epsilon $, but the converse is not true. The functional form of 
$K(q'', q';\tau )$ is generally different from $\tilde{K}(q, q';\epsilon 
)$. The latter can be an approximation of the former. However, the 
short-time propagator has to obey the normalization condition (17). It must 
also satisfy the semi-group property (18) when the exponential 
contributions of ${\cal O}(\epsilon ^{2})$ are ignored. Keeping this 
subtle difference in mind, we use now on the same notation for the short-time 
propagator and the finite-time propagator by dropping the tilde from the 
short-time propagator.

Then we utilize the techniques in harmonic analysis to evaluate the 
path integral on a homogeneous space. The transformation group of the 
homogeneous space may be directly related to a geometrical symmetry 
of the system in question. Or it may be of a dynamical origin.

In order to convert a path integral defined on the homogeneous space ${\cal M}$ 
into one on the group manifold $G$, we restrict the short-time propagator 
$K(q,q';\epsilon)$ to be symmetric under the interchange of two end points, $q 
\rightleftharpoons q'$, and invariant under the action of $g \in G$. 
Namely, 
\begin{equation}
K(q,q';\epsilon )=K(q',q;\epsilon )=K(gq',gq;\epsilon ), ~~~
{\rm for ~all}~~g \in G.
\end{equation}
Then, we introduce for the fixed $q_{a}$ the following function,
\begin{equation}
u_{\epsilon }(g)=K(q_{a}, gq_{a};\epsilon ).
\end{equation}
It is obvious from the properties (19) that  
\begin{equation}
K(q, q';\epsilon )=u_{\epsilon }(g^{-1}g')=u_{\epsilon }(g'^{-1}g)
~~{\rm for}~~q=gq_{a}~,~~ q'=g'q_{a}.
\end{equation}
Furthermore, it is easily to verify that
\begin{equation}
u_{\epsilon }(g)=u_{\epsilon }(h_{1}^{-1}gh_{2})~~
{\rm for~all}~~h_{1},h_{2}\in H.
\end{equation}
Hence, the function $u_{\epsilon }(g)$ is a zonal function \cite{G50,BG62}.

Consequently, we can express Feynman's path integral (16) as the limit 
of a multi-convolution (see Appendix),
\begin{equation}
K(q'',q';\tau )=\lim_{N\to\infty }\int\limits_{G}\d g_{1}\cdots
\int\limits_{G}\d g_{N-1}\,
u_{\epsilon }(g_{0}^{-1}g_{1})\cdots u_{\epsilon }(g_{N-1}^{-1}g_{N}) 
\end{equation}
or 
\begin{equation}
K(q,q_{0};\tau )=\lim_{N\to\infty }\prod_{j=1}^{N}\,*\,u_{\epsilon 
}(g_{j-1}^{-1}g_{j}) 
\end{equation}
where $q_{j}=g_{j}q_{a}$, $j=0,\ldots,N$.
Thus, the path integral in a homogeneous space is reduced to a 
convolution in a group manifold $G$.

Since the short-time propagator $u_{\epsilon }(g)$ is a spherical function which becomes 
constant on the two-sided cosets $HgH$, it can be expanded in terms of 
the zonal spherical functions $D^{l}_{00}(g)$. At this point, 
however, we make an assumption that the group $G$ can be given as 
a direct product $A \otimes B$ of two unimodular groups $A$ and $B$, and 
the isotropy group $H$ of $G$ at $q_{a}$ is a subgroup of $A$. Let $a 
\in A$, $b \in B$ and $g=ab \in G$. Then, the zonal spherical function 
$D^{l}_{00}(g)$ of $G$ may be decomposed in terms of the zonal 
spherical functions $D^{l}_{00}(a)$. As a result, the short-time 
propagator can be expressed as 
\begin{equation}
u_{\epsilon }(g)=\sum_{l\in\Lambda }d_{l
}\,\lambda _{l}(b; \epsilon ) \,D_{00}^{l}(a)
\end{equation} 
where $\Lambda $ stands for the set of all spherical representations (see 
Appendix) and the expansion coefficients are 
\begin{equation}
\lambda _{l}(b; \epsilon )=\int\limits_{A}\d a \, u_{\epsilon }(ab)\,
D^{l}_{00}(a^{-1}).
\end{equation}

Inserting the series expansion (25) into (23) and noticing that $\d g=\d a\d b$, 
we calculate the convolution. 
Use of the orthogonality relation of the zonal spherical functions (see 
Appendix) immediately leads to 
\begin{equation}
K(q'',q';\tau )=\sum_{l\in\Lambda }d_{l}\,K_{l}(b_{0}^{-1}b_{N};\tau )\,
D_{00}^{l}(a_{0}^{-1}a_{N})
\end{equation}
where
\begin{equation}
K_{l}(b_{0}^{-1}b_{N};\tau )=\lim_{N \rightarrow \infty }\,\prod_{j=1}^{N}\, 
* \lambda _{l}(b_{j-1}^{-1}b_{j}; \epsilon ).
\end{equation}
The last convolution will be calculated later for a specific example. 

If, in particular, the subgroup $B$ consists of the unit element $e$ alone, 
that is, if $g=a$, we have
\begin{equation}
K_{l}(e;\tau )=\lim_{N \rightarrow \infty } [\lambda _{l}(\epsilon)]^{N}.
\end{equation}
Since the normalization condition (17) must be satisfied, we obtain 
\begin{equation}
\lim_{\epsilon \rightarrow 0}\lambda _{l}(\epsilon )=1~~{\rm 
for~all}~~ l \in\Lambda .
\end{equation}
This limiting condition allows us to calculate the remaining limit in (29). 
To be more explicit, 
\begin{equation}
\lim_{N\to\infty }\left[\lambda _{l}(\tau /N)\right]^{N}=
\lim_{N\to\infty }\left[1+(\tau /N)\dot{\lambda }_{l}(0)\right]^{N}=
\exp\left\{\tau \dot{\lambda }_{l}(0)\right\}
\end{equation}
where $\dot{\lambda }_{l}(\epsilon )=\d \lambda _{l}(\epsilon )/\d 
\epsilon $. The resulting propagator reads
\begin{equation}
K(q'',q';\tau )=\sum_{l\in\Lambda }^{~}d_{l}\,\exp\left\{\dot{\lambda }_{l}(0)
\tau \right\}\,D^{l}_{00}(g_{0}^{-1}g_{N}).
\end{equation}

In this particular case, since the Hilbert space can be decomposed as ${\cal 
H}=\bigoplus\limits_{l\in\Lambda} {\cal H}^{l}$,
we can make the spectral decomposition of the Hamiltonian,
\begin{equation}
H=\sum_{l\in\Lambda }\sum_{m} \,E_{l}\,|l,m \rangle \langle l,m |
\end{equation}
where $E_{l}$ are the eigenvalues of $H$ and $\{|l,m\rangle\}$ is a complete 
orthonormal basis in ${\cal H}^l$.
Therefore, we have 
\begin{equation}
K(q'',q';\tau )
=\sum_{l,m}\exp\{-(\i/\hbar)E_{l}\tau \} 
\langle q''|l,m \rangle \langle l,m |q'\rangle.
\end{equation}
Comparing this with the result (32) we can identify the spectrum as well as 
the corresponding eigenstates of $H$:
\begin{equation}
E_{l}=\i\hbar\dot{\lambda }_{l}(0)~~,~~~~
\langle q|l,m\rangle=\sqrt{d_{l}}\,\,D^{l}_{m0}(g).
\end{equation}
Although this result is rather obvious,  it provides a simple 
prescription to obtain the correct Hamiltonian associated with a given 
semiclassical short-time propagator. This will be illustrated in the 
example discussed in the following section.

\section{Quantum Mechanics on a Space of Constant Negative Curvature}
Recent interest in quantum mechanics of classically chaotic systems has
revived the study of quantum motion on spaces of constant negative
curvature. For example, the classical free motion on a two-dimensional compact
space of constant negative curvature exhibits chaotic behavior
\cite{G85}. Naturally, it is important to investigate the quantum
aspect of such a motion.

In what follows, we shall study quantum-mechanically a particle of
mass $M$ on a $d$-dimensional non-compact space of constant negative
curvature, which is an integrable system.

The line element of a space having curvature $~K=-1/R^{2}$
is given by \cite{MTW70}
\begin{equation}
\d s^{2}=(1+r^{2}/R^{2})^{-1} \d r^{2}+r^{2}\d \Omega^{2}.
\end{equation}
Here, $R ~( > 0)$ is the ``radius'' of curvature, and
$\d\Omega^{2}$ represents the $(d-1)$-dimension\-al angular part of
the line element, which is identical with that of a flat Euclidean space
$\rz^{d}$. Introducing a new variable $\theta $ by setting $\sinh\theta
=r/R$ ~$(\theta \geq 0)$, the line element can be put into the form
\begin{equation}
\d s^{2}=R^{2}\d\theta ^{2}+R^{2}\sinh^{2}\theta \d\Omega^{2}.
\end{equation}
This geometry can be embedded in a $(d+1)$-dimensional pseudo-Euclidean
space as follows:
\begin{equation}
{\bf x}=R\sinh\theta\, {\bf e}(\Omega), ~~~~x_{d+1}=R\cosh\theta
~~~~\Rightarrow ~~~~ \d s^{2}=\d{\bf x}^{2}-\d x_{d+1}^{2}.
\end{equation}
The vector ${\bf e}(\Omega )$ is a unit vector in the Euclidean subspace
$\rz^{d}$ having polar coordinates $\Omega $. Remember that $\Omega $ stands
short for the $d-1$ angular coordinates.
By this embedding the space ${\cal M}$ can be identified  with the upper sheet of a
time-like subspace of constant "radius" $R$ of a pseudo-Euclidean space with
metric $(+1,\ldots,+1,-1)$.

The transformation group acting transitively on ${\cal M}$ is $SO(d,1)$.
The isotropy group about the origin $\theta =0$ is the compact subgroup
$SO(d)$. This means that the space ${\cal M}$ may be identified with the
quotient space $SO(d,1)/SO(d)$. The variables, $\theta $ and $\Omega $,
indicating the position of the particle on ${\cal M}$, may be identified
with the group parameters. Namely, a group element of $SO(d,1)$ which
transforms the origin $q_{a}=(0,0)$ into a point $q=(\theta , \Omega )$
by $q=gq_{a}$ is parameterized as $g=g(\theta ,\Omega )$. The invariant
Lebesgue measure on ${\cal M}$ is related to the Haar measure of
$SO(d,1)$ (as defined in Ref.\ [\onlinecite{V68}] p.\ 509):
\begin{equation}
\int\limits_{\cal M}\d\theta \d\Omega \,\sinh\theta \,
f\Bigl(q(\theta ,\Omega)\Bigr)=
\frac{2\pi ^{(d+1)/2}}{\Gamma \left(\frac{d+1}{2}\right)}
\int\limits_{SO(d,1)}\d g\,f(gq_{a}).
\end{equation}
This relation is indeed compatible with (A.2) of the appendix.
Note that the Haar measure is only unique up to a multiple constant.

Having set up the geometry of the configuration space, we have to
construct the short-time propagator for the particle in motion. The
Lagrangian for the classical free motion in the above geometry is given
by
\begin{equation}
L=\frac{m}{2}\,\left(\frac{\d s}{\d t}\right)^{2}.
\end{equation}
As Feynman realized already in his 1948 paper, we must recognize the
special role of cartesian coordinates. In the present problem, there is
no cartesian coordinate system. The coordinates closest to the
cartesian ones are those given in (37). In analogy with the
Ozaki-Edward-Gulyaev prescription in formulating the polar coordinate
path integral, we choose \cite{J89b}:
\begin{equation}
S_{\epsilon }(q,q')=\frac{MR^{2}}{\epsilon }\,
(\cosh\Theta -1)+\frac{\hbar^{2}\epsilon }{8MR^{2}}
\end{equation}
where $\Theta $ is the angle of the hyperbolic rotation $g=g(\Theta )$
transforming $q'$ into $q$, $q=gq'$. When analytical continuation in
$\Theta $ is made, this short-time action goes over to the counterpart
for the compact group $SO(d+1)$. There is ambiguity in the choice of
the additive constant. However, this is a common problem for both the
compact and non-compact cases. For a free motion, there is no decisive
criterion, physical or mathematical, for the selection of the constant term.
With the short-time action chosen above, we have the short-time
propagator,
\begin{equation}
K(q,q';\epsilon )=\left(\frac{MR^{2}}{2\pi \i\hbar\epsilon }\right)^{d/2}
\exp\{(\i/\hbar)S_{\epsilon}(q,q')\}.
\end{equation}
The prefactor has been chosen in order to comply with the normalization
condition (17).

In passing from the path integral on the homogeneous space ${\cal M}$ to
the one on the group manifold $G=SO(d,1)$, we have to bear the relation (39) in
mind. Then, we apply the Fourier transformation to the short-time
propagator. Since there are no bound states for the free motion, the
energy spectrum expected are continuous. Therefore, we use the
irreducible unitary representations of $SO(d,1)$ belonging to the
fundamental series characterized by a complex label $l=l(\rho )=-
(d-1)/2 + \i \rho $, $~(\rho \geq 0)$. The integral decomposition of the
short-time propagator is
\begin{equation}
\frac{2\pi ^{(d+1)/2}}{\Gamma (\frac{d+1}{2})}K(q,q';\epsilon )=\int\limits_{0}^{\infty }\d\rho \,d_{l}\,\lambda
_{l}(\epsilon )\,D_{00}^{l}({g'}^{-1}g).
\end{equation}
In this representation, the zonal functions are explicitly given by Gegenbauer functions \cite{V68}
\begin{equation}
D^{l}_{00}(g)=\frac{\Gamma (d-1)\Gamma (l+1)}{\Gamma (l+d-1)}\,
C_{l}^{(d-1)/2}(\cosh\Theta )
\end{equation}
and are related to those of $SO(d+1)$ by analytical continuation in
$\Theta $ and $l$. The constant $d_{l}$ is given by \cite{BJ87}
\begin{equation}
d_{l}=2\,\frac{|\Gamma (\frac{d-1}{2}+\i\rho )|^{2}}{\Gamma (d)|\Gamma
(\i\rho )|^{2}}.
\end{equation}

The Fourier coefficients calculated with (26) take the form \cite{BJ87},
\begin{equation}
\lambda _{l}(\epsilon )=\left(\frac{2MR^{2}}{\pi\i\hbar\epsilon }\right)^{1/2}
\exp\left\{\frac{\i\hbar\epsilon}{8MR^{2}}\right\}\,
K_{\i\rho }(MR^{2}/\i\hbar\epsilon )
\end{equation}
where $K_{\nu }(z)$ is the modified Bessel function of the third kind.
Use of the limiting relation,
\begin{equation}
\lim_{N\rightarrow\infty }\left[\sqrt{\frac{2Nz}{\pi }}\e^{1/(8Nz)}\e^{Nz}
K_{\i\rho }(Nz)\right]^{N}=\exp\{-\rho ^{2}/2z\}
\end{equation}
which is obtained from the asymptotic form of the Bessel function
$$
K_{\i\rho }(Nz)\sim \sqrt{\pi Nz/2}\,\e^{-Nz}\left(1-\frac{\rho ^{2}+1/4}{2Nz}
\right),~~~N\rightarrow\infty ,
$$
enables us to put the propagator into the form \cite{BJ87}
\begin{equation}
K(q'',q';t)=\frac{\Gamma (\frac{d+1}{2})}{2\pi ^{(d+1)/2}}
\int\limits_{0}^{\infty }\d\rho \, d_{l}\,
\exp\left\{-\frac{\i}{\hbar}\,\frac{\hbar^{2}\rho ^{2}}{2MR^{2}}\,t\right\}
D^{l}_{00}(g_{0}^{-1}g_{N}).
\end{equation}
Obviously the spectrum of the Hamiltonian associated with (48) is given by
$E_{\rho }=\hbar^{2}\rho ^{2}/2MR^{2}$ $~(\rho \geq 0)$. This may be compared
with the spectrum of the Laplace-Beltrami operator $\Delta$ on
${\cal M}$ \cite{V68}
which has the spectrum $l(l+d-1)$.
The Hamiltonian which corresponds to our choice (41) of the short-time
action is
\begin{equation}
H=-\frac{\hbar^{2}}{2M}\,\Delta+\frac{(d-1)^{2}}{8MR^{2}}\,\hbar^{2}.
\end{equation}
There have also been other suggestions for short-time actions in the
literature \cite{BJ87,GS87}. However, the present choice (41)
has the consequence that for its corresponding Hamilton operator (49) Huygens'
principle is valid \cite{LP78}.

Although the propagator cannot be given in a simple form, its
Laplace transform, the energy-dependent Green function, can also be given in closed
form \cite{Comment}.
The propagator (48) can be written, if $d$ is odd,
in the form \cite{J89b}
\begin{equation}
K(q'',q';\tau )=R\left(\frac{M}{2\pi \i\hbar \tau }\right)^{1/2}
\left(\frac{-1}{2\pi\sinh\Theta }\,\frac{\partial}{\partial\Theta }
\right)^{(d-1)/2}\exp\{ (\i/\hbar)S_{cl}(\Theta,\tau )\}
\end{equation}
where $S_{cl}(\Theta,\tau )=(MR^{2}\Theta ^{2}/2\tau )$ is the classical action and
$\Theta $ is the hyperbolic angle between the two positions $q'$ and
$q''$.
For $d$ even, it takes a similar but different form,
\begin{equation}
\begin{array}{ll}
\displaystyle
K(q'',q';\tau )=&\displaystyle\sqrt{2}R^{2}
\left(\frac{M}{2\pi\i\hbar\tau }\right)^{3/2}
\left(\frac{-1}{2\pi\sinh\Theta}
\frac{\partial}{\partial\Theta}\right)^{(d-1)/2}\\[2mm]
&\times\displaystyle
\int\limits_{\Theta }^{\infty }\d z\,
\frac{z\exp\{(\i MR^{2}/\hbar 2\tau )z^{2}\}}{\sqrt{\cosh z-\cosh\Theta }}.
\end{array}
\end{equation}

\section{The Harmonic Oscillator}
The harmonic oscillator is so common that it can hardly be an attractive 
object. Historically, it has been the object that Feynman's path 
integral can handle. As is mentioned earlier, the hydrogen atom is too 
difficult to path-integrate. However, thanks to several techniques 
developed in the past ten years, it has become possible to reduce many 
non-quadratic path integrals including the one for the hydrogen atom 
into the path integral for the harmonic oscillator.  By the harmonic 
oscillator, here, we don't mean a linear harmonic oscillator. We mean an 
isotropic harmonic oscillator in higher dimensions. A path integral of 
the confluent hypergeometric class is reducible to the path integral for 
the radial harmonic oscillator in an inverse-square potential. 
Performing path integration in polar coordinates even for 
the harmonic oscillator is not trivial. In fact, the path integration in polar 
coordinates was done already in 1967 by brute force \cite{PI69}. It 
was only after the work of B\"ohm and Junker \cite{BJ87} when the study 
of the link between the radial propagator and the matrix element of the 
$SU(1,1)$ began. Because of the spherical symmetry, it is expected that 
the propagator is decomposable in terms of the spherical functions. 
However, it is somewhat surprising that there is a beautiful underlying 
structure in the radial path integral that is connected to the 
non-compact group $SU(1,1)$.

\subsection{The Dynamical Group}
First we examine the group structure of an isotropic harmonic oscillator 
in $\rz^{d}$, having mass $M$ and spring constant $M\omega ^{2}$. The 
Hamiltonian is given by 
\begin{equation}
H=\frac{1}{2M}\,{\bf p}^{2}+\frac{1}{2}M\omega ^{2}{\bf x}^{2}.
\end{equation}

The dynamical group of the three-dimensional isotropic harmonic 
oscillator has been identified with the symplectic group $Sp(6)$ by 
Moshinski and Quesne \cite{MQ71}. A subgroup $Sp(2)\otimes SO(3)$ of 
$Sp(6)$ has been chosen as the basis for the separation in spherical 
polar coordinates. 
The group $SO(3)$ is associated with the rotational symmetry of the 
oscillator, and the symplectic group $Sp(2)$, which is isomorphic with 
$SU(1,1)$, is the spectrum-generating group. Generalizing this to the 
$d$-dimensional case, we use the group $SU(1,1) \otimes SO(d)$ to 
separate the system in polar coordinates. 

Let us realize the algebra of $SU(1,1)$ by 
\begin{equation}
\begin{array}{rcl}
J_{1}&=&\displaystyle
-\frac{1}{4M\hbar\omega}\left({\bf p}^{2}-M^{2}\omega^{2}{\bf 
x}^{2}\right),~~~~
J_{2}=\displaystyle
-\frac{1}{4\hbar}\left({\bf x}\cdot {\bf p} + {\bf p}\cdot {\bf 
x}\right),\\[3mm]
J_{3}&=&\displaystyle\frac{1}{4M \hbar \omega }\left({\bf p}^{2} +
M^{2}\omega ^{2}{\bf x}^{2}\right).
\end{array}
\end{equation}
Namely, these operators satisfy 
\begin{equation}
\left[ J_{1},J_{2}\right]=-\i J_{3},~~~
\left[ J_{2},J_{3}\right]=\i J_{1},~~~
\left[ J_{3},J_{1}\right]=\i J_{2}.
\end{equation}
The Casimir operator is ${\bf J}^{2}=-J_{1}^{2}-J_{2}^{2}+J_{3}^{2}$. 
On the standard orthonormal basis ${|J, \mu \rangle}$, 
\begin{equation}
{\bf J}^{2}|J,\mu \rangle =J(J+1)|J,\mu  \rangle~~,~~~~
J_{3}|J,\mu  \rangle=\mu |J,\mu  \rangle.
\end{equation}
The unitary irreducible representations of $SU(1,1)$ are labeled by the 
number $J$. There are two discrete series denoted by $D_{J}^{+}$ and 
$D_{J}^{-}$ and two continuous series denoted by $C_{J}$ and 
$E_{J}$ \cite{B47}.

In the present realization, the Hamiltonian is proportional to $J_{3}$ as
\begin{equation}
H=2\hbar \omega J_{3}.
\end{equation}
Since $J_{3}$ is positive-definite by construction, the eigenvalues of 
$J_{3}$ must be positive. Therefore, the set of the operators (53) 
provides a realization of the representation $D_{J}^{+}$ of $SU(1,1)$ 
for which $\mu = -J + n$ $~(n \in \nz _{0})$ \cite{BF66}. 

It can be verified that the Casimir operator ${\bf J}^{2}$ 
calculated with (53) is related to the Casimir invariant ${\bf L}^{2}$ 
of $SO(d)$ as 
\begin{equation}
{\bf J}^{2}=\frac{1}{4\hbar^{2}\,}{\bf L}^{2} + \frac{1}{16}\, d(d-4). 
\end{equation}
Here ${\bf L}$ is the angular momentum in $\rz^{d}$ whose components are  
$L_{ik}=x_{i}p_{k} - x_{k}p_{i} ~(i,k=1,2,\dots,d)$.
The Casimir invariant of $SO(d)$,
\begin{equation}
{\bf L}^{2}=\frac{1}{2}\sum_{i,k=1}^{d} L_{ik}L_{ik},
\end{equation}
has the spectrum $\hbar ^{2} l(l + d - 2) ~(l \in \nz_{0})$. Therefore, we have
\begin{equation}
J(J+1)=\frac{1}{4}l(l+ d -2) + \frac{1}{16}d(d - 4). 
\end{equation}

Thus, the subspace of the Hilbert space ${\cal H}={\cal L}^{2}(\rz^{d})$ with 
a fixed angular momentum $l$ carries also the irreducible representation space 
of $D_{J(l)}^{+}$ with 
\begin{equation}
J(l)= - \frac{1}{2}l - \frac{1}{4}d.
\end{equation}
The Hilbert space ${\cal H}$ may be decomposed into an orthogonal sum of 
subspaces ${\cal H}^{l}$ with $l \in \nz _{0}$. Each subspace is a 
product of the $SU(1,1)$ representation $D_{J}^{+}$ and the $SO(d)$ 
representation denoted by $D^{l}$. Namely, 
\begin{equation}
{\cal L}^{2}(\rz^{d})={\cal L}^{2}(\rz^{+})\otimes {\cal L}^{2}(S^{d-1})=
\bigoplus\limits_{l=0}^{\infty }\left(D^{+}_{J(l)}\otimes D^{l}\right).
\end{equation}
The basis in ${\cal H}={\cal L}^{2}(\rz^{d})$ is given as a 
tensor product of the discrete basis $\{|J(l),\mu \rangle \}$ in 
$D^{+}_{J(l)}$ and the basis $\{|l, m\rangle\}$ in $D^{l}$, that 
is,
\begin{equation}
|\mu,l, m\rangle=|J(l),\mu \rangle\otimes|l, m\rangle .
\end{equation}
Obviously, these are the eigenstates of the Hamiltonian (56):
\begin{equation}
H|\mu,l,m\rangle=\left(2\hbar\omega J_{3}\otimes 
{\bf 1}\right)\,|\mu,l,m\rangle=2\hbar\omega \mu |\mu,l, m
\rangle . 
\end{equation}
Using $\mu = - J(l) + \nu  =\nu + l/2 + d/4 = n/2 + d/4 $, we obtain the 
expected result,   
\begin{equation}
E_{n}=\hbar\omega (n + d/2)~~,~~~~n = 2\nu + l \in\nz_{0}~.
\end{equation}

Here we have to note that the evolution operator $\exp\{-
(\i/\hbar)H\tau \}$ which is now identified with $\exp\{-2\i\omega 
\tau J_{3}\}$ is an element of $SU(1,1)$. The third generator $J_{3}$ 
of $SU(1,1)$ acts as the generator of the one-parameter group of time 
evolution. The group $SU(1,1)$ is literally the dynamical group of the 
harmonic oscillator.                   

The orthonormal set $\{|J,\mu  \rangle\} $ used above forms a discrete basis, 
which diagonalizes the compact operator $J_{3}$. 
Although the discrete basis is widely used, the use of a continuous 
basis which diagonalizes a non-compact operator is often overlooked.
There have been extensive studies of continuous bases \cite{BP68,LN70,MR73}. 
For instance, $K_{+}=J_{1}+J_{3}$ is a non-compact operator. The basis 
which makes $K_{+}$ diagonal is a continuous basis:
\begin{equation}
K_{+}|J,\eta  \rangle =\eta |J,\eta  \rangle 
\end{equation}
where the eigenvalue $\eta $ is a continuous variable. In the present 
realization, it happens to be  
\begin{equation}
K_{+}= \frac{M\omega }{2\hbar}{\bf x}^{2}.
\end{equation}
Obviously, the eigenvalue of $K_{+}$ is 
\begin{equation}
\eta = \alpha r^{2} 
\end{equation}
where $\alpha =M\omega /(2\hbar)$ and $r=|{\bf x}|.$  The corresponding 
eigenstates in ${\cal H}$ are given by 
\begin{equation}
|\eta ,l, m\rangle=|J(l),\eta  \rangle\otimes|l, m\rangle .
\end{equation}

In parallel to Wigner's $d$-function, 
$d_{mn}^{l}(\theta )=\langle l,m | \e^{-\i\theta J_{2}}|l,m\rangle $ 
defined for the compact group $SU(2)$, Bargmann defined the functions, 
$b_{mn}^{l}(\theta )=\langle l,m |\e^{-\i\theta J_{2}}|l,m\rangle $ 
on the discrete basis of $SU(1,1)$. In the case of $SU(2)$, all the 
group generators are compact, whereas $J_{3}$ of $SU(1,1)$ 
is the only compact operator and $J_{2}$ is non-compact.  The function
defined by Bargmann is a matrix element of a non-compact member of 
the $SU(1,1)$ group on the discrete basis. It is certainly interesting 
to define a function which is a matrix element of a member of the 
maximal compact subgroup of $SU(1,1)$ on the continuous basis that 
diagonalizes $K_{+}$. Namely, 
\begin{equation}
v_{\eta \eta '}^{J}(\theta )=\langle J,\eta |\exp\{-\i\theta J_{3}\}|J,
\eta '\rangle 
\end{equation}
where $0 < \theta < 2\pi $. 
In the unitary irreducible representation of the discrete series 
$D_{J}^{+}$, it has been shown by Lindblad and Nagel
that the $v$-function takes the following explicit form \cite{LN70,I92}, 
\begin{equation}
v_{\eta \eta '}^{J}(2\varphi)=
\frac{1}{\i\sin\varphi }\,\exp\left\{\i(\eta +\eta ')\cot\varphi 
\right\}\,\I_{-2J-1}\left(\frac{2\sqrt{\eta \eta '}}{\i\sin\varphi 
}\right),
\end{equation}
where $\I_{\nu }(z)$ is the modified Bessel function of the first kind and 
$0< \varphi  <\pi $. This function turns out to be the core function of 
the radial propagator of the harmonic oscillator.

\subsection{The propagator}
According to the relation (56), the propagator as a matrix element of the
evolution operator can be put in the form,
\begin{equation}
K({\bf x}'', {\bf x}'; \tau )=\langle {\bf x}'' \mid \exp\{-2\i\omega \tau
J_{3}\} \mid {\bf x}' \rangle .
\end{equation}
Using the basis (68) which diagonalizes the non-compact operator $K_{+}$,
we can express (71) as
\begin{equation}
K({\bf x}'', {\bf x}'; \tau )=\sum_{l,m} \int\limits_{\rz ^{+}}\!\d\eta ''
\int\limits_{\rz ^{+}}\!\d\eta ' \,
\langle {\bf x}'' |\eta '', l, m \rangle \langle J(l), \eta ''
|\e^{-2\i\omega\tau J_{3}}| J(l), \eta ' \rangle  \langle \eta ', l, m
| {\bf x}' \rangle .
\end{equation}
Let ${\bf x}=r {\bf u}$ where $r=|{\bf x}|$ and ${\bf u}={\bf x}/r$,
and let also ${\bf u}=h {\bf u}_{a}$ where $h \in SO(d)$ and ${\bf
u}_{a}$ is a fixed point on $S^{d-1}$. Then we can write the position
states as $|{\bf x}'\rangle = |r', h'{\bf u}_{a} \rangle $ and $|{\bf
x}''\rangle = |r'', h''{\bf u}_{a} \rangle $, where $h', h'' \in SO(d)$.
The propagator (72) can now be expressed in the form,
\begin{equation}
K({\bf x}'', {\bf x}'; \tau )=\frac{\Gamma (d/2)}{2\pi ^{d/2}}
\sum_{l=0}^{\infty }K_{l}(r'', r' ; \tau )\, d_{l}\, D^{l}_{00}(h''^{-
1}h')
\end{equation}
with
\begin{equation}
K_{l}(r'', r';\tau )= \int\limits_{\rz ^{+}} \d \eta ''
\int\limits_{\rz ^{+}} \d \eta ' \,
\langle r'' \mid \eta '' \rangle \langle J(l), \eta ''
\mid \e^{-2\i\omega \tau J_{3}} \mid J(l), \eta ' \rangle  \langle
\eta ' \mid r' \rangle .
\end{equation}
The dimension of the representation $D^{l}$ is given by
\begin{equation}
d_{l}=(2l+d-2)\,\frac{(l + d-3)!}{l! \,(d-2)!},
\end{equation}
and the zonal spherical functions are expressed in terms of Gegenbauer
polynomials
\begin{equation}
D^{l}_{00}(h''^{-1}h')=\frac{l!\, \Gamma
(d-2)}{\Gamma (d + l -2)}\,C_{l}^{(d-2)/2}({\bf u}''\cdot {\bf u}').
\end{equation}
Note that the area of $S^{d-1}=SO(d)/SO(d-1)$ is $2\pi ^{d/2}/\Gamma
(d/2)$.

Since $\d\eta =2\alpha r\,\d r$, and since
the completeness relations of the sets,  $\{ | r \rangle \}$ and $\{ |
J(l),\eta \rangle \}$, are given by
\[
\int\limits_{\rz ^{+}}\d r\,r^{d-1}\, |r \rangle \langle r |  = 1, ~~~~~
\int\limits_{\rz ^{+}}\d\eta \, |J(l), \eta \rangle \langle J(l), \eta |  = 1.
\]
we have
\begin{equation}
\langle r \mid J(l),\eta \rangle = \sqrt{2\alpha }
r^{-(d-2)/2}\,\delta ( \eta - \alpha r^{2}), ~~~~~\alpha =M\omega /(2\hbar).
\end{equation}
Making use of this relation in (74), we can immediately obtain the radial propagator
in the form,
\begin{equation}
\begin{array}{ll}
K_{l}(r'', r';\tau )&=\displaystyle
\frac{M\omega }{\hbar}(r'r'')^{-(d-2)/2}\,
\langle J(l),\eta'' | \exp\{-2\i\omega \tau J_{3}\} | J(l), \eta ' \rangle \\[2mm]
&\displaystyle=
\frac{M\omega }{\hbar}(r'r'')^{-(d-2)/2}\,
v^{J(l)}_{\eta '' \eta '}(2 \omega \tau )
\end{array}
\end{equation}
where $\eta '= (M\omega /2\hbar)r'^{2}$ and
$\eta''=(M\omega/2\hbar)r''^{2}$.\\

\subsection{Path Integration for the Propagator}
Next, let us make an explicit calculation of Feynman's path
integral for the oscillator:
\begin{equation}
K({\bf x}'', {\bf x}'; \tau )=\lim_{N\rightarrow\infty }
\int\limits_{\rz^{d}}\prod_{j=1}^{N-1} \d{\bf x}_{j}
\prod_{j=1}^{N} K({\bf x}_{j}, {\bf x}_{j-1}; \epsilon ) \,=
\lim_{N\rightarrow\infty }
\prod_{j=1}^{N}\,* K({\bf x}_{j}, {\bf x}_{j-1}; \epsilon ).
\end{equation}

The Lagrangian corresponding to the
Hamiltonian (52) is
\begin{equation}
L=\frac{1}{2}M\dot{\bf x}^{2} - \frac{1}{2}M\omega ^{2}\,{\bf x}^{2}.
\end{equation}
The short-time propagator of the harmonic oscillator in $ \rz ^{d}$ is
given by
\begin{equation}
K({\bf x}, {\bf x}'; \epsilon )=
\left(\frac{M}{2\pi\i\hbar\epsilon}\right)^{d/2}
\exp\left\{\frac{\i}{\hbar}\left[\frac{M}{2\epsilon}({\bf x} -
{\bf x}')^{2}-
\frac{M}{4}\omega^{2}\epsilon ({\bf x}^{2}+{\bf x'}^{2})\right]\right\}.
\end{equation}
The prefactor of the exponential function is chosen so as to meet the
normalization condition (3).
Since the kinematic factor $\exp[(\i M/2\hbar\epsilon )({\bf x}-{\bf x}')^{2}]$
is invariant under the group of motion in $\rz ^{d}$, it can
be expanded in terms of the zonal spherical functions of the rotation
group $SO(d)$ or the zonal spherical functions of the Euclidean group
$E(d)$. If we put it into the form,
\begin{equation}
\exp\left[\frac{\i M}{2\hbar\epsilon }\left({\bf x}-{\bf x}'\right)^{2}\right]
=
\exp\left[\frac{\i M}{2\hbar\epsilon }\left(r^{2} + r'^{2}\right)\right]\,
\exp\left[-\frac{\i M}{2\hbar\epsilon }\,{\bf x}\cdot {\bf x}'\right],
\end{equation}
the second factor is a zonal function with respect to $SO(d)$.
The expansion of the second factor in terms of the zonal spherical function
$D_{00}^{l}(h)$ of $SO(d)$ turns out to be the well-known Gegenbauer expansion,
\[
\exp( z \,{\bf u}\cdot {\bf u}')=(2/z)^{\lambda }\Gamma (\lambda )
\sum_{l=0}^{\infty } (l + \lambda )\,\I_{l + \lambda }(z)
\,C_{l}^{\lambda }({\bf u}\cdot {\bf u}'),
\]
with $z=(M/2\i\hbar\epsilon )$ and $\lambda =(d-2)/2$. Consequently, the
short-time propagator can be expanded as
\begin{equation}
K({\bf x}, {\bf x}'; \epsilon )=\frac{\Gamma (d/2)}{2\pi ^{d/2}}
\sum_{l=0}^{\infty }K_{l}(r, r' ; \epsilon )\, d_{l}\, D^{l}_{00}(h^{-
1}h')
\end{equation}
with the short-time radial propagator,
\begin{equation}
\begin{array}{ll}
K_{l}(r, r'; \epsilon )=&\displaystyle
-2\i(\alpha/\epsilon ) \,(rr')^{-(d-2)/2}
\exp\left[\i(\alpha /\omega \epsilon )\left(1 - \frac{1}{2}\omega
^{2}\epsilon ^{2}\right)(r^{2}+r'^{2})\right] \\[2mm]
&\displaystyle
\times \I_{l+(d-2)/2}\left(\frac{\alpha }{\i\omega \epsilon }rr'\right).
\end{array}
\end{equation}
For convenience, let us define the following function,
\begin{equation}
v_{\lambda }(\eta ,\eta ';\varphi )=-\i\csc \,\varphi\, \exp[\i(\eta + \eta
')\cot \varphi ]\,\I_{\lambda }\left(-2\i\sqrt{\eta \eta '}\,\csc\,\varphi
\right).
\end{equation}
If we use Weber's integral formula,
\begin{equation}
\int\limits_{0}^{\infty }\d r \, r\,\exp(\i\beta r^{2})\,\I_{\lambda }(-\i ar)\,
\I_{\lambda }(-\i br) =
\frac{\i}{2\beta }\exp\left[-\frac{\i}{4\beta }(a^{2} + b^{2})\right]\,
\I_{\lambda }\left(-\i\frac{ab}{2\beta }\right),
\end{equation}
valid for Re $\beta > 0$ and Re $\lambda > -1$, then we can verify the
recurrence convolution,
\begin{equation}
v_{\lambda } * v_{\lambda }(\eta '', \eta ';\varphi ''+\varphi )=
\int\limits_{0}^{\infty }\d\eta\, v_{\lambda }(\eta '', \eta ; \varphi '')\,v_{\lambda
}(\eta , \eta '; \varphi ) = v_{\lambda }(\eta '', \eta ';
\varphi '' + \varphi ).
\end{equation}
This convolution formula can be extended to the multi-convolution formula,
\begin{equation}
v_{\lambda }(\eta _{N}, \eta _{0}; \varphi ) = \prod_{j=1}^{N}\, *
v_{\lambda }(\eta _{j}, \eta _{j-1}; \varphi _{j}),
\end{equation}
where
\begin{equation}
\varphi = \sum_{j=1}^{N}\,\varphi _{j}.
\end{equation}

Now we can utilize this convolution formula to calculate the path
integral as the multi-convolution of the short-time propagator. The
multi-convolution of the $N$ short-time propagators is given by
\begin{equation}
K({\bf x}_{N}, {\bf x}_{0}; N\epsilon )=\prod_{j=1}^{N} * K({\bf x}_{j},
{\bf x}_{j-1}; \epsilon ) =\frac{\Gamma (d/2)}{2\pi ^{d/2}}
\sum_{l=0}^{\infty }K_{l}(r_{N}, r_{0} ; N\epsilon )\, d_{l}\,
D^{l}_{00}(h_{N}^{-1}h_{0}),
\end{equation}
with
\begin{equation}
K_{l}(r_{N}, r_{0} ; N\epsilon )=\prod_{j=1}^{N} \,* K_{l}(r_{j}, r_{j-1}).
\end{equation}
If we let
\begin{equation}
\sin \varphi _{j}= \omega \epsilon , ~~~~~ \cos \varphi _{j}= 1 -
\frac{1}{2}\omega ^{2} \epsilon ^{2} + {\cal O}(\epsilon ^{4}),
\end{equation}
the short-time radial function may be approximated as
\begin{equation}
K_{l}(r, r' ; \epsilon ) = 2\alpha (rr')^{-(d-2)/2}\,v_{l + (d-
2)/2}(\alpha r^{2}, \alpha r'^{2}; \varphi _{j}).
\end{equation}
Therefore, the multi-convolution formula of the $v$-function helps to
evaluate the multi-convolution of the radial propagator. In the limit $N
\rightarrow \infty $ (i.e., $\epsilon \rightarrow 0$),
\[
\varphi = \lim_{N \rightarrow \infty } N {\rm arcsin}(\omega \epsilon ) = \omega \tau .
\]
As a result, we obtain the radial propagator for a finite time interval
$\tau $,
\begin{equation}
K_{l}(r'', r' ; \tau ) = 2\alpha (r''r')^{-(d-2)/2}\,v_{l + (d-
2)/2}(\alpha r''^{2}, \alpha r'^{2}; \omega \tau ),
\end{equation}
or
\begin{equation}
\begin{array}{ll}
K_{l}(r'', r' ; \tau ) =&\displaystyle - 2\i\alpha (r''r')^{-(d-2)/2}\,\csc
(\omega \tau )\exp\left[\frac{\i M\omega }{2\hbar} (r'^{2} + r''^{2}) \cot(\omega
\tau )\right] \\[2mm]
&\displaystyle\times
\I_{l+(d-2)/2}\left(\frac{M\omega}{\i\hbar}r'r''\csc(\omega \tau )\right).
\end{array}
\end{equation}

Comparison of (78) and (95) leads us to the relation,
\begin{equation}
\langle J,\eta | \e^{-2\i\varphi J_{3}} |J,\eta '\rangle =
-\i \csc \,\varphi\, \exp[\i(\eta + \eta ')\cot \varphi ]\,
\I_{-2J-1}\left(-2\i\sqrt{\eta \eta '}\,\csc\,\varphi \right)
\end{equation}
which coincides with the result (70) obtained by Lindblad and Nagel
\cite{LN70}. In this way, we can also determine the matrix element on
the continuous basis by path integration.

Shapiro and Vilenkin \cite{SV86} expressed Weber's integral formula (86)
in terms of the matrix element in question. Here, conversely, we have used
Weber's formula to determine the matrix element.\\

\section{Epilogue}
Fourier analysis is well-known as a tool for finding the heat kernel.
Then it is not surprising that harmonic analysis works
well in calculating various kernels of the Sch\"odinger equation. The
path integral itself is known as a tool of finding the Schr\"odinger
kernel or the Feynman kernel or the propagator in non-relativistic
quantum mechanics. What is surprising is that path integrals and harmonic
analysis can mix very well. In the preceding sections, we have
demonstrated how the techniques of harmonic analysis can be utilized in
path integration in nontrivial ways.

We have picked up only two examples, but the same or similar techniques
used in those examples have been applied to various other problems. As
generalized Fourier expansions may be applied to ({\it i}) functions on
the homogeneous space ${\cal M}=G/H$, ({\it ii}) zonal functions defined
on spheres $S=HgH$, and ({\it iii}) central functions on the group
manifold $G$, so can the propagators be treated differently, depending
on their structure. In the first example, we have expanded the
propagator in terms of zonal functions. However, in dealing with a
particle on a three-dimensional sphere, we may apply the character
expansion. This is because $S^{3}$ can be identified with the group
manifold of $SU(2)$. In this case, the group $SU(2)$ is directly
related to the rotational symmetry of the background space. The
one-dimensional P\"oschl-Teller oscillator, for instance, has no spherical
symmetry, but can be solved on the dynamical group manifold $SU(2)$.
Even though the physical natures of the group $SU(2)$ for these two
cases are different, the structures of the two path integrals are
basically the same. Both belong to the hypergeometric class. Path
integrals of the hypergeometric class solved by harmonic analysis
include the free motion on a space of constant positive curvature
$S^{d}= SO(d+1)/SO(d)$, the free motion on the Euclidean space $\rz^{d}$
by means of the zonal function of the Euclidean group, and the rotations
about the origin $\rz^{d}=E^{d}/SO(d)$. For details, see refs.
[\onlinecite{J89b,BJ89,J89a}].

For the second example, we have used the matrix elements on the
continuous basis. The radial path integral belongs to the confluent
hypergeometric class. A number of problems have also been solved with the
help of the radial path integral of this type. For a review of the path
integrals of $SU(2)$ and $SU(1,1)$, we refer to ref.\ [\onlinecite{I92}].

In concluding, we may say that these group theoretical methods together with
other techniques (see also [\onlinecite{I92}]) have made Feynman's path integral
as accessible as Schr\"odinger's equation.

\section*{Acknowledgement}
We would like to thank Raj Wilson for his invitation to this stimulating
conference from which we greatly benefited.

%\input{ctexasa.tex}
%\inputfile{txa.tex}
\setcounter{equation}{0}
\def\theequation{A.\arabic{equation}}

\section*{Appendix: Rudiments of Harmonic Analysis on Homogeneous Spaces}
This Appendix reviews the rudiments of harmonic analysis on homogeneous 
spaces in a way pertinent to path integral calculus. For details, see 
refs.\ \cite{V68} and \cite{BR80}.

\subsection*{Transformation Group and Homogeneous Space:}
First, we consider a transformation group $G$ of a space ${\cal M}$. A
transformation $g$ of ${\cal M}$ is a one-to-one map of ${\cal M}$ onto 
itself. If $q$ is a point of ${\cal M}$, then the transform of $q$ by 
$g$, denoted by $q'=gq$, also belongs to ${\cal M}$. A set $G$ of such 
transformations is a transformation group of ${\cal M}$ if the inverse 
$g^{-1}$ transforms $q'$ back into $q$ by $q=g^{-1}q'$ and if 
$(g_{1}g_{2})q=g_{1}(g_{2}q)$ for any two transformations, $g_{1}$ and 
$g_{2}$, of $G$. We assume that $G$ acts transitively on ${\cal M}$. In other 
words, for every pair $(q', q)$ of ${\cal M}$, there is a transformation $g 
\in G$ such that $q'=gq$. 
We also assume that $G$ is effective on ${\cal M}$, that is, 
for every group element $g$ not equal to the identity element there exists
a $q\in{\cal M}$ such that $gq\neq q$.
Then, ${\cal M}$ is a homogeneous space with 
respect to the transitive group $G$. A subgroup $H$ of $G$ whose action 
leaves a point $q_{a}$ of ${\cal M}$ fixed is called the {\it isotropy 
group} of $G$ at $q_{a}$. Since any point $q$ of ${\cal 
M}$ can be reached from the fixed point $q_{a}$ by a group action $g 
\in G$, we have $q=gq_{a}=ghq_{a}$ where $h \in H$. Thus, there is one-to-one 
correspondence between the homogeneous space ${\cal M}$ and the coset 
space $G/H$. Hence ${\cal M}$ is identified with $G/H$, that is, 
${\cal M}=G/H$. 

Suppose a fixed point $q$ of ${\cal M}$ is taken to a point $q'=hq$ of 
${\cal M}$ under the action $h$ of the isotropy group $H$ at $q_{a}$. 
The set $S=\{q'\in hq; \,h \in H \}$ is called {\it the sphere} centered at 
$q_{a}$, passing through $q$. Since $q$ corresponds to the left coset 
$gH$, the sphere $S$ can be viewed as the two-sided coset $HgH$ with $g$ 
fixed. 

Next, we assume that the transformation group $G$ is locally compact and 
unimodular. Then, $G$ has a unique (up to a 
multiple constant) invariant Haar measure for all integrable functions 
$f: ~G \mapsto \cz $, denoted by d$g$: 
\begin{equation}
\int\limits_{G}\d g\,f(g)=
\int\limits_{G}\d g\,f(g'g)=
\int\limits_{G}\d g\,f(gg')=
\int\limits_{G}\d g\,f(g^{-1})
\end{equation}
where $g' \in G$. The Haar measure d$g$ 
induces a $G$-invariant measure d$q$ on ${\cal M}$ \cite{BR80}:
\begin{equation}
\int\limits_{{\cal M}}\d q \,f(q)=\int\limits_{G}\d g \,f(gq_{a}) 
\end{equation}
where $q_{a}$ is a fixed point in ${\cal M}$ and $f: ~{\cal M} \mapsto 
\cz $ is an integrable function 
on ${\cal M}$. The $G$-invariance of the measure on ${\cal M}$, 
\begin{equation}
\int\limits_{{\cal M}}\d q \,f(gq)=\int\limits_{{\cal M}}\d q \,f(q),
\end{equation}
is a direct consequence of the invariance of the Haar measure.

\subsection*{Spherical Representations:} 
Let us denote by $D^{l}$ a unitary irreducible representation of 
the group $G$ in an invariant subspace 
${\cal H}^{l}\subset{\cal L}^{2}({\cal M})$. 
The representation $D^{l}(g)$ is called a {\it spherical} 
representation (or a representation of {\it class 1}) of $G$ if ${\cal 
H}^{l }$ has non-null vectors $|\psi _{\alpha } \rangle$ which are 
invariant under transformations of the subgroup $H$ \cite{V68}, 
$D^{l }(h)|\psi _{\alpha } \rangle=|\psi _{\alpha } \rangle $ for 
all $h\in H$. Let the subgroup $H$ be {\it massive} \cite{V68}, that is, 
let there be only one such vector $|\psi _{0}\rangle$ in ${\cal 
H}^{l }$. Then a function $\psi ^{l}(g)$ defined on $G$ by 
\begin{equation}
\psi ^{l }(g)=\langle \psi \mid D^{l }(g) \mid \psi _{0} \rangle 
\end{equation}
is called a {\it spherical function} of $D^{l}(g)$. It is obvious that 
$\psi ^{l}(gh)=\psi ^{l}(g)$. Thus, a spherical function $\psi ^{l}(g)$ 
of $D^{l}(g)$ is constant on a left coset $gH$. Since ${\cal M}$ is 
identified with the space of the left cosets $gH$, $\psi ^{l}(g)$ can be 
regarded as a function on the homogeneous space ${\cal M}=G/H$. 

Let $\{|e_{m} \rangle \}$ with $ \langle e_{m} \mid e_{n} \rangle = 
\delta _{mn}$ ~$(m,n=0,1,2,\dots, d_{l}-1)~$ be an orthonormal basis with
$|e_{0} \rangle = | \psi _{0} \rangle $
in the $d_{l}$-dimen\-sional subspace ${\cal H}^{l}$. 
The matrix elements of $D^{l}(g)$ for $g \in G$ on this basis are 
given by 
\begin{equation}
D^{l}_{mn}(g)=\langle e_m \mid D^{l}(g) \mid e_n \rangle .
\end{equation}
The unitary property implies $D^{l~\ast }_{nm}(g)=D^{l
}_{mn}(g^{-1})$. It is also easy to show that 
\begin{equation}
D^{l}_{mn}(g_{1}g_{2})=\sum_{k} D^{l}_{mk}(g_{1})D^{l
}_{kn}(g_{2}).
\end{equation}
Certainly the matrix elements
\begin{equation}
D^{l}_{m 0}(g)=\langle e_{m}\mid D^{l}(g) \mid e_{0} \rangle 
\end{equation}
are spherical functions of $D^{l}(g)$ on $G$, which are more 
specifically called {\it associated spherical functions}. They satisfy 
$D^{l}_{m 0}(gh) = D^{l}_{m 0}(g)$ for any $h \in H$.  A special 
case of associated spherical functions is 
\begin{equation}
D^{l}_{0 0}(g)=\langle e_{0}\mid D^{l}(g) \mid e_{0} \rangle ,
\end{equation}
which is called a {\it zonal spherical function}. Naturally, the zonal 
spherical function is constant on the two-sided coset $HgH$, that is, 
for any $h, h' \in H$, 
\begin{equation}
D^{l}_{00}(h^{-1}gh')=D^{l}_{00}(g).
\end{equation}

For instance, if $G=SO(3)$ and if $H=SO(2)$ about the north pole 
$q_{a}=(0,0,1)$ of ${\cal M}=S^{2}$, then the well-known spherical 
harmonics $Y^{m}_{l}(\theta ,\phi )$ are associated spherical functions 
of irreducible unitary representations of $SO(3)$, and the Legendre 
polynomials $P_{l}(\cos \theta )$ are the zonal spherical functions. The 
latter has a constant value along a circle (i.e., $\theta =$ constant) on 
$S^{2}$. 

Let us add a few more examples. The zonal spherical functions on an 
$d$-dimen\-sional Euclidean space $E^{d}$, the unit sphere $S^{d-
1}=SO(d)/SO(d-1)$ in $d$ dimensions, and the $(d-1)$-dimensional 
Loba\v{c}evski\u{i} space are given, respectively, by 
\[ 
\begin{array}{l} 
D^{k}_{00}(r)=2^{\nu }\Gamma (\nu +1 )\,(kr)^{-\nu }\,J_{\nu }(kr), 
~~~~(0 <  k, r < \infty ), \vspace{2mm} \\ 
D^{\ell }_{00}(\theta )=\frac{\Gamma (\ell +1)\Gamma (2\nu )}{\Gamma 
(\ell +2\nu )}C^{\nu }_{\ell }(\cos \theta ), ~~~~(\ell 
=0,1,2,...;~0\leq \theta \leq\pi ), \vspace{2mm} \\ 
D^{\rho }_{00}(t)=2^{\nu -1/2}\Gamma (\nu + \frac{1}{2})(\sinh t)^{-\nu 
+1/2}\,P^{\frac{1}{2}-\nu }_{-\frac{1}{2}+\i\rho }(\cosh t), ~~~~
(0< \rho ,t < \infty ), 
\end{array} 
\] 
where $2\nu =d-2$. In the above, $J_{\nu }(z), C^{\nu }_{l}(z)$, and 
$P^{\nu }_{\mu }(z)$ are the Bessel function, the Gegenbauer polynomials, 
and the associated Legendre function, respectively.

The character $\chi ^{l}(g)$ of a finite dimensional irreducible 
unitary representation $D^{l}$ is defined by 
\begin{equation}
\chi ^{l}(g) = {\rm Tr} D^{l}(g) = \sum_{m}\,D^{l
}_{mm}(g).         
\end{equation}
In particular, $\chi ^{l}(e) = d_{l} =$ dim ${\cal H}^{l}$ 
where $e$ is the unit element of $G$. For an infinite dimensional 
representation, the character cannot be defined in the way it is defined 
for the finite dimensional case. The character of an infinite 
dimensional representation is defined as a distribution. 

\subsection*{Fourier Expansions:}
In the case of compact groups, the harmonic analysis in the Hilbert 
space ${\cal H}={\cal L}^{2}(G)$ is based on the Peter-Weyl theorem. 
The theorem tells us that the matrix elements, 
$\sqrt{d_{l}}\, D^{l}_{mn}(g)$, form a complete orthonormal 
set on ${\cal H}$ in regard to a normalized invariant Haar measure $\d g$,
satisfying the orthogonality relation, 
\begin{equation}
\int\limits_{G} \d g\, D^{l}_{mn}(g)\, D^{l'\,*}_{m'n'}(g)=d_{l}^{-
1}\,\delta _{ll'}\delta _{mm'}\delta _{nn'}.
\end{equation}
An arbitrary function $f(g) \in {\cal H}$ may be expanded in the form,
\begin{equation}
f(g)=\sum_{l, m,n}\,d_{l}\,\hat{f}^{l}_{mn}\, 
D^{l}_{mn}(g),
\end{equation}
where 
\begin{equation}
\hat{f}^{l}_{mn}=\int\limits_{G}\d g\,f(g)\,D^{l\,*}_{mn}(g). 
\end{equation}

For a non-compact group $G$, the spectra of some invariant operators of 
$G$ are continuous, and the eigenstates should be understood as 
distributions. Therefore, we have to deal with the so-called {Gel'fand 
triplet} $\Phi \subset {\cal H} \subset \Phi '$. Here, $\Phi $ is 
a certain nuclear space of smooth functions in ${\cal H}$, and $\Phi '$ 
is the dual space of $\Phi $. The orthogonality relation (A.11), which 
holds only in special cases, should be broadly interpreted. For 
instance, the Kronecker delta $\delta _{ll'}$ may be 
replaced by the delta function $\delta (l - l')$, and the 
``dimension'' $d_{l}$ may be defined by the relation (A.11) itself. 
The Fourier expansion formula (A.12) may be expressed in the form,
\begin{equation}
f(g)=\sum_{l\in \Lambda } 
\,d_{l}\,\mbox{Tr}\Bigr(\hat{f}^{l}D^{l}(g)\Bigl),
\end{equation}
where $\Lambda $ is the set of all inequivalent unitary irreducible 
representations to which $D^{l}$ belongs. The summation in the 
above expression must be replaced by an appropriate Lebesgue-Stieltjes 
integral when $ l $ is continuous. 
  
The zonal spherical functions (A.9) satisfy the relations, 
\begin{equation}
\int\limits_{G}\d g\,D_{00}^{l}(g)D_{00}^{l'}(g^{-
1})=\frac{1}{d_l}\,\delta_{ll'}
\end{equation}
and 
\begin{equation}
\int\limits_{G}\d g_{j}\,D_{00}^{l}(g_{j-1}^{-1}g_{j})
D_{00}^{l'}(g_{j}^{-1}g_{j+1})=\frac{1}{d_l}\,\delta _{ll'}
D_{00}^{l}(g_{j-1}^{-1}g_{j+1}),
\end{equation}
both of which readily follow from the orthogonality relation (A.11). 
Any function $f(g)$ constant on two-sided cosets $HgH$, that is, 
satisfying $f(h^{-1}gh)=f(g)$ for $h \in H$ and $g \in G$, can be 
expanded in terms of the zonal spherical function $D^{l}_{00}(g)$ 
as 
\begin{equation}
f(g)=\sum_{l\in \Lambda }\,d_{l}\,\hat{f}^{l
}\,D^{l}_{00}(g),
\end{equation}
where the ``Fourier'' coefficients are given by
\begin{equation}
\hat{f}^{l}=\int\limits_{G} \d g\,f(g) \,D^{l}_{00}(g^{-
1}).
\end{equation}

It is evident from the orthogonality relation (A.11) that the character 
function (A.10) satisfies 
\begin{equation}
\int\limits_{G}\d g\, \chi ^{l\,\ast }(g)\,\chi ^{l'}(g) 
= \delta _{ll'}, 
\end{equation}
and 
\begin{equation}
\int\limits_{G}\d g_{2} \, \chi^{l}(g_{3}g_{2}^{-1}) \chi ^{l'}(g_{2} g_{1}^{-
1})= d_{l}^{-1} \,\delta_{ll'} \,\chi 
^{l}(g_{3} g_{1}^{-1}). 
\end{equation}
A function $f(g)$ on $G$, satisfying $f(g_{1}^{-1}g_{2}g_{1})=f(g_{2})$ 
for any $g_{1}, g_{2} \in G$, is called a {\it central function}. 
Evidently, the character $\chi ^{l}(g)$ is a central function on 
$G$. Any central function on $G$ can be decomposed by means of the 
character function:
\begin{equation}
f(g)=\sum_{l\in \Lambda }\,d_{l}\,\hat{f}^{l}\,\chi ^{l}(g),
\end{equation}
with 
\begin{equation}
\hat{f}^{l}=d_{l}^{-1}
\int\limits_{G}\d g\, f(g)\,\chi ^{l}(g^{-1}).
\end{equation}
For a compact group, the presence of the dimensional constant 
$d_{l}$ in the series expansion seems redundant. However, if $G$ 
is non-compact, and if $ l $ takes a continuous value, then the 
replacement of the sum by an integral necessarily brings the factor 
$d_{l}$ as part of the Jacobian. Therefore, for a unified 
treatment of the discrete and continuous cases, it is convenient to keep 
$d_{l}$ explicitly in (A.21) and (A.22).

%\noindent Convolutions:\\
\subsection*{Convolutions:}
The convolution $f_{1} * f_{2}(g)$ of two integrable functions 
$f_{1}(g)$ and $f_{2}(g)$ on a locally compact unimodular group $G$ is 
defined by 
\begin{equation}
f_{1} * f_{2}(g) = \int\limits_{G}\d g_{1}\, f_{1}(gg_{1}^{-1})f_{2}(g_{1}), 
\end{equation}
which is also integrable on $G$. The operation of convolution is 
associative; 
\begin{equation}
f_{1}*(f_{2}*f_{3})=(f_{1}*f_{2})*f_{3}.
\end{equation}
It is sometime convenient to use the following short hand notation 
for a multi-convolution of $N$ functions,
\begin{equation} 
(\cdots ((f_{1} * f_{2}) * f_{3}) * \cdots) * f_{N} = \prod_{j=1}^{N}\, 
* f_{j}. 
\end{equation} 
It is noteworthy that if two square integrable functions $f_{1}(g)$ and 
$f_{2}(g)$ are expanded as 
\begin{equation}
f_{1}(g) = \sum_{l}\,d_{l}\,\sum_{m,n}\,a^{l
}_{mn}\,D^{l}_{mn}(g),
\end{equation}
\begin{equation}
f_{2}(g) = \sum_{l}\,d_{l}\,\sum_{m,n}\,b^{l
}_{mn}\,D^{l}_{mn}(g),
\end{equation}
then the Fourier transform $c^{l}_{mn}$ of their convolution 
$f_{1} * f_{2}(g)$ is given by 
\begin{equation}
c^{l}_{mn}=\sum_{k}\,a^{l}_{mk}b^{l}_{kn}.
\end{equation}

%\begin{thebibliography}{99}
%\input{texasref.tex}

%\end{thebibliography}


\begin{thebibliography}{10}
\bibitem{Comment}Grosche and Steiner \cite{GS87} were the first to
realize that the Green function can be given in closed form, but their
result was incorrect. The correct form of the Green function is given in
ref.\ \cite{J89b}.
\end{thebibliography}
\end{document}